# Tunable Magnetism and Half-Metallicity in Hole-doped Monolayer GaSe

Ting Cao, Zhenglu Li, and Steven G. Louie*
*Department of Physics, University of California at Berkeley, Berkeley, California 94720, USA*
*and Material Sciences Division, Lawrence Berkeley National Laboratory, 1 Cyclotron Road, Berkeley, California 94720, USA*

**Abstract:**
We find, through first-principles calculations, that hole doping induces a ferromagnetic phase transition in monolayer GaSe. Upon increasing hole density, the average spin magnetic moment per carrier increases and reaches a plateau near 1.0 $\mu_B$/carrier in a range of $3\times10^{13}/\text{cm}^2 - 1\times10^{14}/\text{cm}^2$ with the system in a half-metal state before the moment starts to descend abruptly. The predicted magnetism originates from an exchange splitting of electronic states at the top of the valence band where the density of states exhibits a sharp van Hove singularity in this quasi-two-dimensional system.

The study of atomically thin quasi-two-dimensional (2D) crystals has become one of the most rapidly developing areas of condensed matter physics owing to the rich phenomena in and the tremendous promise of applications [1, 2] of these materials. The pursuit of controlled magnetism in 2D crystals in particular has been a persisting goal in this area. In graphene, the formation of magnetic moments and their ordering are theoretically predicted by introducing adatoms, defects, or edges to the system [3-6]. However, experimental measurements of such magnetic moments and resultant magnetic order often yield ambiguous results [7-12]. Even in its realization, the conducting electrons would only become spin polarized by coupling to localized magnetic moments, and they suffer from significant scatterings by the imperfections introduced. Similar proposals have been put forward for monolayer transition metal dichalcogenides [13], but no experimental evidence of formation of magnetic orders at the single layer level has been reported so far.

Recently a new class of 2D crystals, the atomically thin metal monochalcogenides, has attracted much attention [14-22]. In their bulk form, metal monochalcogenides (such as GaS, GaSe, InSe, and GaTe) are layered materials composed of covalently bonded metal and chalcogenide atoms in nanometer thick layers that are coupled to neighboring layers by weak ionic and van der Waals forces. Like graphene, these materials may be thinned down to form 2D crystals. By using a chemical vapor deposition (CVD) method, monolayer GaSe has moreover been successfully synthesized [17, 21, 22]. Large-area crystalline GaSe monolayer can be grown on various substrates, and exhibits high photoresponse and p-type transport behavior [21, 22]. The 2D unit cell of a GaSe monolayer is composed of two Ga atoms and two Se atoms. The two neighboring Ga atoms are vertically stacked, and are sandwiched between two planes of Se atoms [Fig. 1(a)]. The in-plane projected geometry of GaSe monolayer crystal forms a honeycomb crystal structure, in which the projected Ga atoms and Se atoms occupy the two triangular sublattices respectively [Fig. 1(b)].

Although bulk GaSe has a direct band gap according to theoretical calculations [16, 21, 23], a direct to indirect band gap transition occurs as the number of layers is decreased below some critical value, a trend opposite to that of the transition metal dichalcogenides. Specifically, as the number of layers decreases to below 7, calculations show that the conduction band minimum remains at the Brillouin zone center ($\Gamma$), but the energy of the states of the highest occupied band near $\Gamma$ is suppressed leading to the valence band maximum (VBM) shifting progressively away from $\Gamma$ to nearby k-points, resulting in a Mexican-hat-like energy surface around the zone center [21]. At the monolayer limit, the energy difference between the VBM and the highest valence band state at $\Gamma$ is maximized, with the Mexican-hat-like landscape extending over a significant fraction of the Brillouin zone. This unusual character of the band structure gives rise in a high density of states (DOS) and an almost one-dimensional (1D) like van Hove singularity near the VBM.

A large DOS at or near the Fermi level ($E_F$) of a system, in general, would lead to instabilities and transitions to different phases such as magnetism, superconductivity, and other phenomena [3, 24, 25]. Also, a sharp feature in the DOS near $E_F$ is often indicative of a material with a large thermoelectric Seebeck coefficient [26, 27]. Here, we report a first-principles theoretical investigation of magnetism in monolayer GaSe as $E_F$ is tuned, via hole doping, through the van Hove singularity in the DOS. We show that there is Stoner-type magnetic instability with hole doping leading to a half-metallic ferromagnetic ground state. Ferromagnetism is predicted to appear at a significant range of carrier densities, with the density of magnetic moment and spin-polarized carrier tunable by changing the doping level.

We performed first-principles calculations using density functional theory (DFT) in the local density approximation (LDA) for the exchange-correlation function as implemented in the Quantum Espresso package [28]. We employed fully relativistic norm conserving pseudopotentials, with a plane-wave energy cutoff for the wavefunctions of 100 Ry. We set the lattice constant to 3.75 Å, corresponding to its experimental measured value for the monolayer and the bulk [21, 29]. The atomic positions in the unit cell were fully relaxed until the force on each atom was smaller than 0.01 eV/Å. Since the band structure and the DOS near the VBM are central to the present study, we have also performed quasiparticle band structure calculations within the *ab initio* GW approach [30] as implemented in the BerkeleyGW package

[31].

In the calculations, carrier density was tuned by changing the total number of electrons in the unit cell, with a compensating jellium background of opposite charge added. For various doping conditions, it is found that a minimum number of 8,100 (i.e. 90×90) k-points was required in sampling the Brillouin zone to converge the magnetic moment, since the magnetic instability depends sensitively on the sampling of the DOS near $E_F$.

The calculated DFT-LDA Kohn-Sham band structure at zero doping (p = 0) is shown in Fig. 2(a) which has an indirect Kohn-Sham band gap of 2.0 eV. At the top valence band, as discussed above, the energy band in the vicinity of the Γ point exhibits a Mexican-hat-like dispersion within a large fraction of the first Brillouin zone. Moving from the Γ point along different in-plane directions, the energy maxima are almost degenerate. For example, at p = 0, the energy maximum along the Γ−K direction is only 0.01 eV higher than that along the Γ−M direction.

A Mexican-hat-like dispersion in two dimensions should give rise to a sharp van Hove singularity in the DOS, resembling that of a band edge in 1D. However, as seen in Fig. 2(b), its divergence behavior in monolayer GaSe is modified by an anisotropic dispersion and spin-orbit splitting which break the cylindrical symmetry of the bands: the DOS is a step function at the VBM and quickly peaks at the energy of -0.013 eV below the VBM. This peak in the DOS originates from saddle critical points of the energy surface for the lower spin-orbit split bands, which are of opposite spin character for the two directions along the Γ−K and Γ−K' directions. A partial density of states (PDOS) analysis shows that the DOS near the VBM is mainly composed of Se 4p orbitals, with a small contribution from Ga 4p and 4s orbitals [Fig. 2(b)]. Quite remarkably, from the point of view of PDOS on each Se atom, it is greater than 2/(eV-atom) for a range of energy near the VBM. This value is comparable to the DOS at $E_F$ of the 3d magnetic transition metals – iron, cobalt, and nickel.

We confirm that the features of the energy dispersion of the top valence bands are robust against many-electron self-energy effects by calculating the quasiparticle band structure. At the $G_0W_0$ level, we obtain an indirect quasiparticle bandgap of 3.7 eV, which is 1.7 eV larger than the LDA Kohn-Sham gap. Despite such a large self-energy correction, the dispersion of the GW band structure of the top valence band almost overlaps with the LDA one [Fig. 2(b)]. A detailed analysis around the Γ point shows that the GW band structure even has slightly less dispersion [Fig. 2(a), inset], which increases the DOS of the system further as compared to the LDA result.

Although intrinsic monolayer GaSe is nonmagnetic, our relativistic DFT-LDA calculations show that it spontaneously develops a ferromagnetic ground state even at a small amount of hole doping. Figure 3 shows the calculated electron spin magnetic moment/carrier (i.e., $\sum_{nk \in hole} -\langle nk|m|nk\rangle / \sum_{nk \in hole}\langle nk|nk\rangle$, where $m$ is the spin magnetic moment operator and $|nk\rangle$ the Bloch states) and the spin-polarization energy/carrier (i.e., the total energy difference between the nonmagnetic and ferromagnetic phase normalized by the number of carriers) as a function of carrier density, for the case with the total electron spin magnetic moment of the system constrained in the out-of-plane direction (along the plane normal). We find an averaged magnetic moment of 0.4 $\mu_B$/carrier at the lowest carrier density considered in our calculations (p = $8\times10^{12}$/cm$^2$), with a very small spin-polarization energy (< 0.1 meV/carrier). Larger spin magnetic moment/carrier develops upon increasing carrier density. At p = $3\times10^{13}$/cm$^2$, the magnetic moment saturates at nearly 1.0 $\mu_B$/carrier. Provided p < $1.0\times10^{14}$/cm$^2$, the magnetic moment/carrier remains at this high value even though the spin carrier density increases by a factor of 3. After this density, further increment of carrier density reduces the magnetic moment/carrier within DFT-LDA. For p > $1.3\times10^{14}$/cm$^2$, monolayer GaSe returns to a nonmagnetic state.

In contrast to a nearly constant magnetic moment/carrier above certain carrier density, the spin-polarization energy has a strong dependence on the doping level. Within DFT-LDA, the spin-polarization energy increases monotonically to 3 meV/carrier at p = $7\times10^{13}$/cm$^2$, and then without having a plateau region decreases monotonically back to 0 at p = $1.3\times10^{14}$/cm$^2$.

In our calculations, hole-doped monolayer GaSe exhibits very weak magnetic anisotropy energy. The calculated energy difference between the out-of-plane spin polarized state and the in-plane spin polarized state is smaller than 0.3 meV/carrier in all carrier densities considered. The dependence of the anisotropy energy on carrier density is shown also in Fig. 3. At both small and large carrier densities, the out-of-plane spin polarization orientation is favored over the in-plane spin polarization orientation, whereas in the intermediate carrier densities, the in-plane spin polarization is favored. The evolution of the magnetic moment/carrier as a function of carrier density for the in-plane spin polarized ferromagnetic state is very similar to that of the out-of-plane spin polarized ferromagnetic state shown in Fig. 3.

Other possible magnetic phases have also been considered in our calculations, including antiferromagnetic states and several spin density waves. We find that all the other phases considered are energetically unfavorable compared with the ferromagnetic state.

The emergent magnetism upon hole doping owes its origin to additional exchange splitting of the bands of the two spin types near $E_F$ by the spontaneously broken of time-reversal symmetry. In Fig. 4, we show the evolutions of the "spin up" band and the "spin down" band at different carrier densities for the state constrained with an out-of-plane spin polarization. The spin moment of the Bloch eigenstates is determined by calculating the expectation value of the electron spin magnetic moment operator along the out-of-plane direction using the spinor wavefunctions. The magnitude of the out-of-plane spin moment of each Bloch eigenstate generally differs from a unit $\mu_B$ due to spin-orbit coupling in this material.

With increasing hole density, a greater exchange splitting of the two spin types appears in the ferromagnetic phase [Fig. 4(a)-(i)]. Due to the large DOS near the

VBM, the exchange fields are very strong, leading to a spontaneously broken of time reversal symmetry in the doped cases shown. A comparison can be made with the strength of the Zeeman splitting caused by an external magnetic field. For example, at p = 7×10$^{13}$/cm$^2$, the splitting at the valence band top along the Γ−K direction is 30 meV larger than that of the undoped case. This energy corresponds to an effective Zeeman splitting from an external magnetic field of about 260 T, if the orbital magnetic contribution is ignored and a g-factor of 2 is assumed.

As another consequence of the strong exchange splitting, hole doped monolayer GaSe in the ferromagnetic state exhibits a half metallic band structure, i.e., the states at $E_F$ are completely from a band of one spin type. In our calculations, at carrier density from 3×10$^{13}$/cm$^2$ to 1.0×10$^{14}$/cm$^2$, the Fermi surface only intersects the "spin up" band [Fig. 4(b)-(d)]. This half metallic behavior allows fully polarized spin transport. In addition to the half metallic behavior, we find a change in the topology of the Fermi surface as the density increases. The shape of the Fermi surface goes from having three pockets to that of a ring, between 3×10$^{13}$/cm$^2$ and 7×10$^{13}$/cm$^2$ carrier densities, respectively.

A key factor to tunable magnetism in monolayer GaSe is controllability of carrier density at relatively high doping levels. For this purpose, atomically thin materials offer a great advantage compared to bulk materials. Because of their thinness, doping carrier densities have been achieved to the order of 10$^{14}$/cm$^2$ in graphene by ion liquid gating [32, 33] and to 10$^{13}$/cm$^2$ in transition metal dichalcogenides monolayers by back-gate gating [34, 35]. We therefore expect that tunable hole doping, to the level discussed in this work, can also be readily achieved for monolayer GaSe in terms of currently available gating techniques. Introduction of p-type dopants and defects may also be a viable means towards achieving the predicted magnetic state. From calculations on substitutional doping with a portion of Se atoms replaced by As atoms in a large supercell, we find that ferromagnetism and similar magnetic properties arise at doping levels discussed in the above calculations with electrostatic doping.

The predicted spontaneous magnetization upon hole doping should be detectable in monolayer GaSe by various methods. In transport measurements, the resistance in the ferromagnetic state is spin-dependent for single-domain samples. Direct detections of the spin polarization employing spin-polarized scan probes such as spin-polarized scanning tunneling microscope (SP-STM) or optical measurements such as the magneto-optic Kerr effect can also give evidence of magnetization. In addition, as seen in Fig. 1(b), monolayer GaSe does not have inversion symmetry. The Berry curvature is thus generally non-vanishing in the 2D Brillouin zone in the undoped case [36, 37]. This property, together with the doping-dependent Fermi surface topology and band splitting, may give rise to anomalous Hall and spin Hall currents in this interesting 2D system.

Moreover, as bulk GaSe is a weakly coupled layered material, the interlayer electronic coupling strength is weak. Consequently, multilayer GaSe with different interlayer stacking configurations or intercalated bulk GaSe samples may offer tunable magnetic properties similar to its monolayer form. We would also like to point out that GaSe is but one of the many layered metal monochalcogenides. Materials such as monolayer GaS and GaTe all have similar crystal and electronic structures [19]. The variety of material in this family offers a great opportunity to explore magnetic phenomena in 2D.

In conclusion, we find through first-principles calculations that hole doping induces tunable ferromagnetism and half metallicity in monolayer GaSe. For a range of hole carrier density, the averaged electron spin magnetic moment can be as large as close to 1.0 $\mu_B$/carrier. We demonstrate that this itinerant magnetism originates from an exchange-field splitting of the electronic states near the VBM, where the DOS for this 2D material is very high and exhibits a sharp van Hove singularity. Our findings not only reveal for the first time possibility of doping induced tunable magnetism in quasi-2D semiconductors, but also open up an opportunity to realize spintronics at the atomically thin single-layer level, in which controlled spin moment and transport may be achieved by electrostatic gating.

**Acknowledgement:**

This research was supported by the Theory Program at the Lawrence Berkeley National Lab through the Office of Basic Energy Sciences, U.S. Department of Energy under Contract No. DE-AC02-05CH11231 which provided the GW calculations, and by the National Science Foundation under Grant No. DMR10-1006184 which provided for the DFT study of doping induced magnetism. This research used resources of the National Energy Research Scientific Computing Center, which is supported by the Office of Science of the U.S. Department of Energy.

**Figures and Captions:**

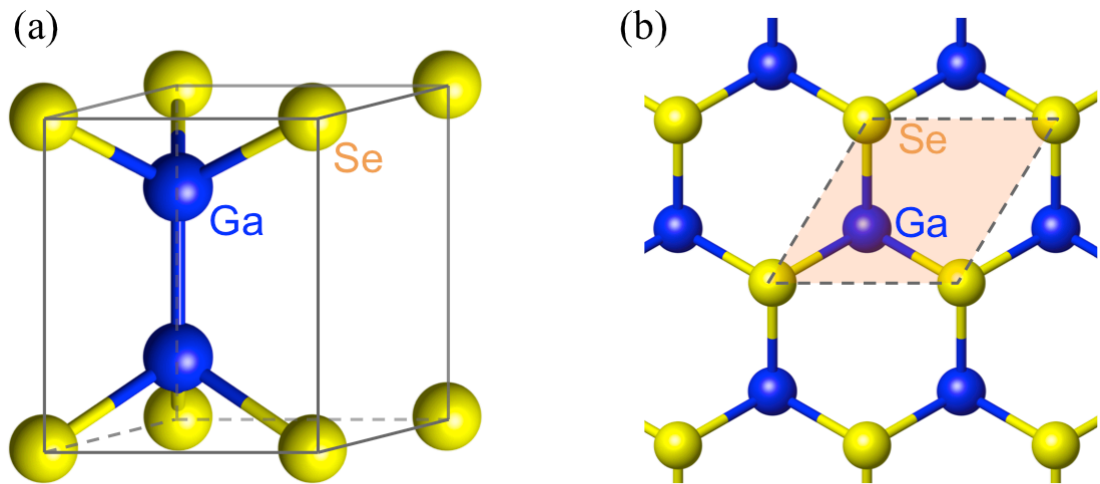

FIG .1. The crystal structure of monolayer GaSe. (a) A side view of the unit cell of GaSe monolayer. (b) A top view of monolayer GaSe crystal. The shaded area shows a unit cell, which consists of two vertically stacked Ga atoms and two vertically stacked Se atoms.

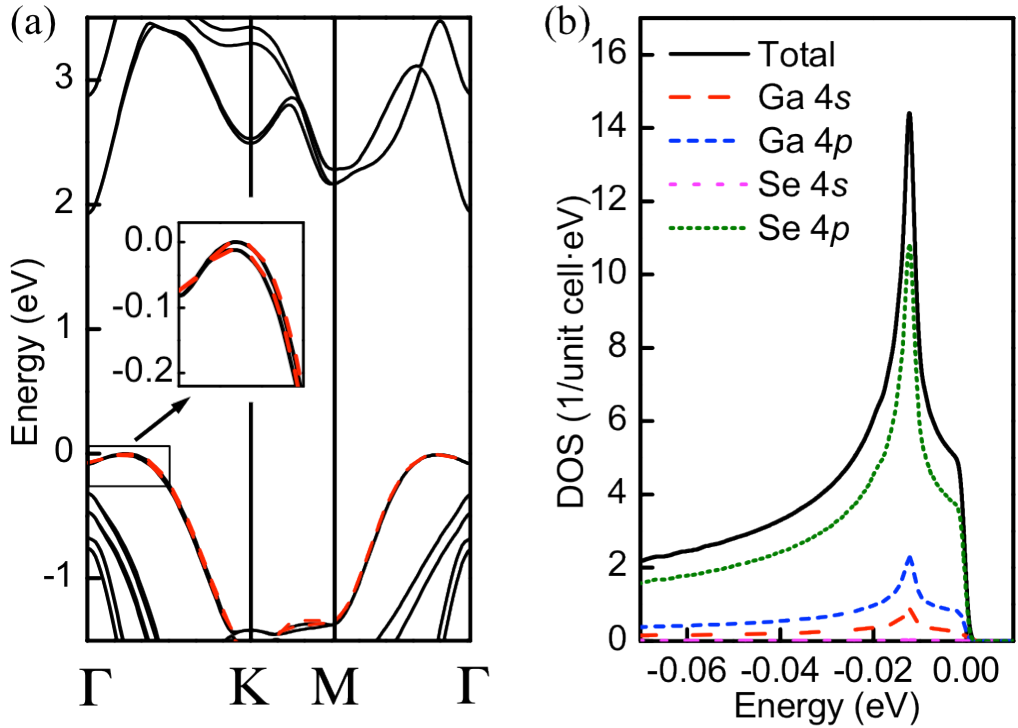

FIG. 2. Electronic structure of undoped monolayer GaSe. (a) Relativistic band structure of GaSe along high symmetry directions. The black solid lines are calculated within density functional theory with the LDA functional. The red dashed lines are the top valence bands calculated by the *ab initio* GW method. The inset shows the band structure zoomed in near the Γ point. (b) Total and partial density of states near the top of the valence band. The total density of states DOS (black solid curve) is in units of states/(eV-unit cell), with a unit cell consisting of two Ga and two Se atoms. The partial density of states (colored curves with different line types) denotes contributions from the different atomic orbitals. A Gaussian broadening of 0.001eV is used.

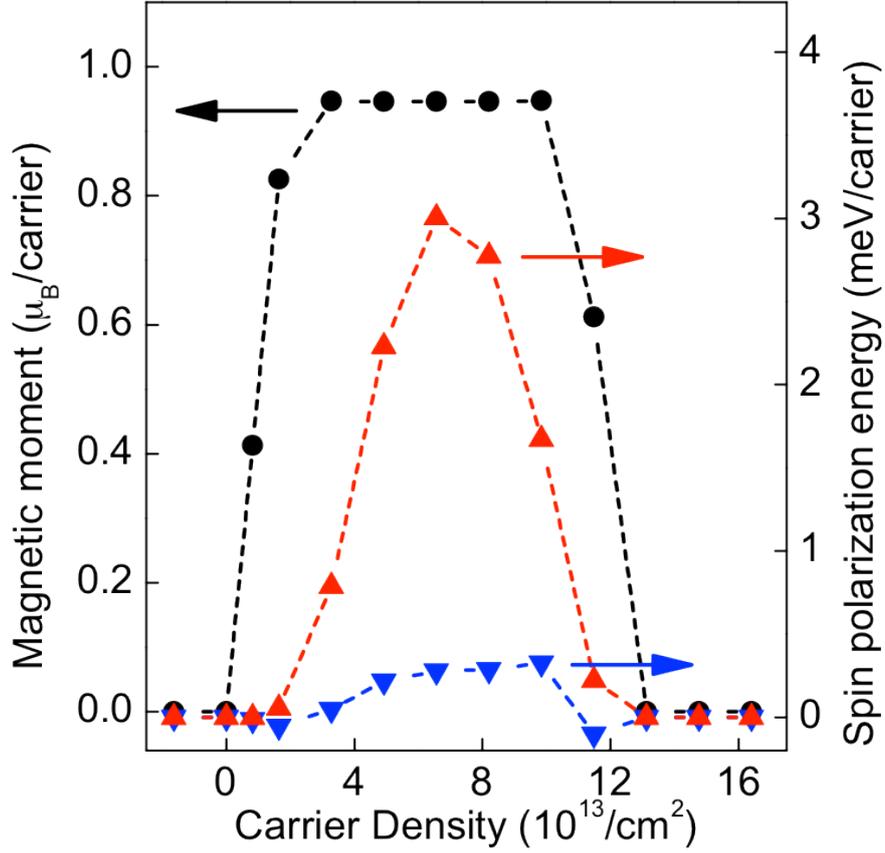

FIG. 3. Carrier density dependent spin magnetic moment/carrier and spin-polarization energy/carrier in the out-of-plane spin polarized ferromagnetic state. The black circles and red triangles denote the magnetic moment/carrier and spin polarization energy/carrier, respectively. The blue inverted triangles denote the magnetic anisotropy energy, defined as the energy difference between the state with out-of-plane polarization orientation and that with in-plane polarization orientation. The dashed lines are guides to the eye.

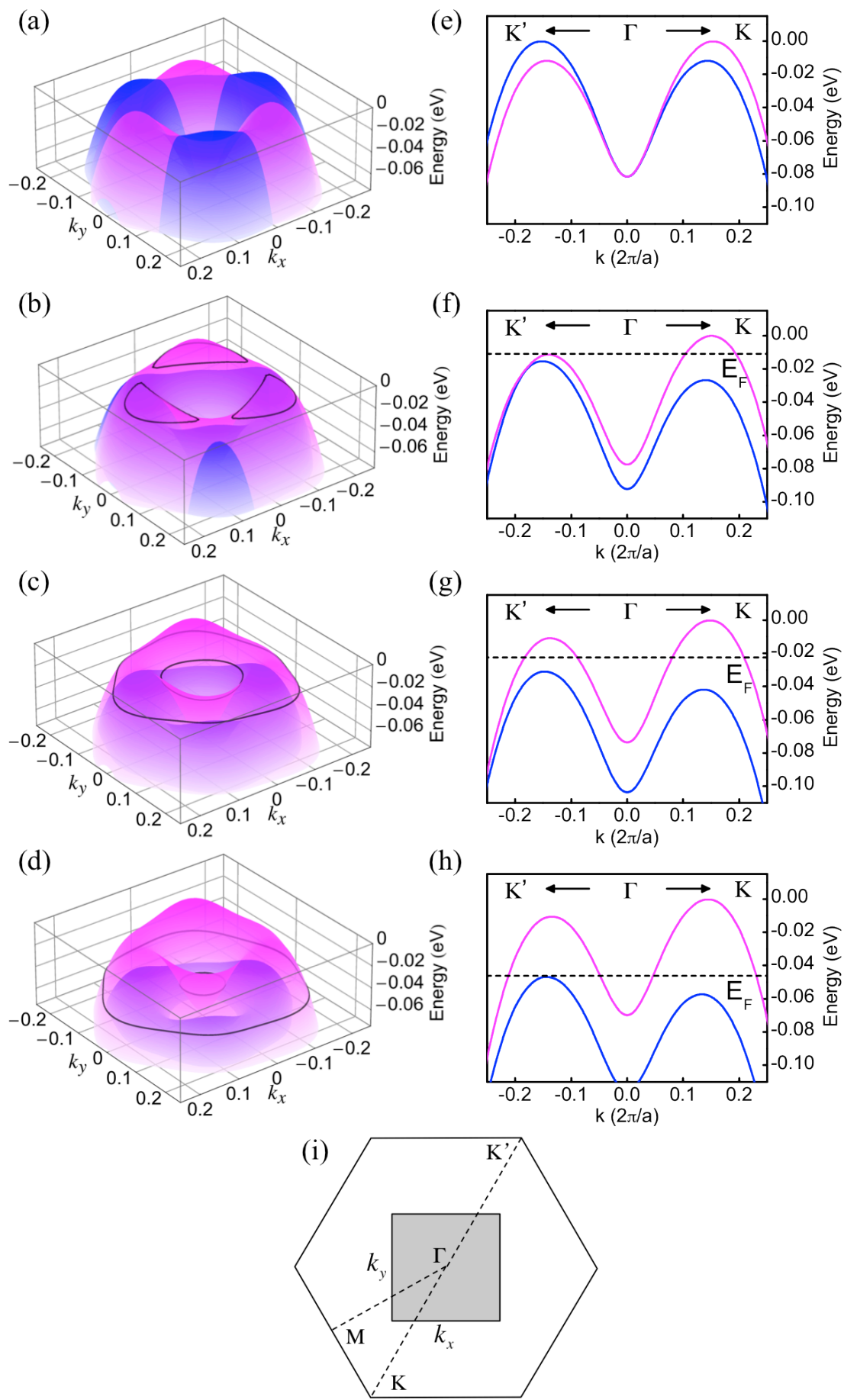

FIG. 4. Carrier density dependent band structure of monolayer GaSe in the out-of-plane spin polarized ferromagnetic state. The purple and blue colors represent "spin up" and "spin down" electronic states, respectively. (a)-(d) 3D plot of the band structures at carrier density of 0, $3\times10^{13}$, $7\times10^{13}$, and $1\times10^{14}$/cm$^2$, respectively. The black curve denotes the shape of the Fermi surface. (e)-(h) Band structures along high symmetry directions at carrier density of 0, $3\times10^{13}$, $7\times10^{13}$, and $1\times10^{14}$/cm$^2$, respectively. The dashed line denotes the Fermi level. (i) The shaded area shows the two-dimensional reciprocal space in which figures (a)-(d) are plotted.

*sglouie@berkeley.edu